# Exact monochromatic finite-power beam solutions of Maxwell's equations in vacuum

**R**OBERT **W**. **H**ELLWARTH,* **L**IUBING **H**UANG

*Department of Electrical Engineering, University of Southern California, Los Angeles, CA 90089*
*Corresponding author: hellwart@usc.edu*

We derived and discussed new exact, monochromatic, finite-power, beam-like solutions of Maxwell's equations in vacuum. We derived energy density flux, total power of the solution, and compared its effective beam area with that of the Gaussian beam solution.

## 1. Introduction.

In this paper we will derive and discuss various exact complex-number beam-like solutions of Maxwell's equations in vacuum; the real and imaginary parts of each of the six complex components of the electric **E** and magnetic **B** vectors of a particular solution comprise two separate independent real-number solutions. Each complex-number solution will be monochromatic, i.e., has its entire dependence on time $t$ contained in a factor $\exp(-i\omega t)$ with $\omega$ a real positive constant. The spatial ($xyz$) variations of each solution clearly exhibit an axis along some arbitrary direction $\hat{z}$. We show that each real solution transmits a definite finite total energy flux $P$ (which we call the beam power) through every ($xy$) plane that is perpendicular to $\hat{z}$. We call these monochromatic finite-power beam solutions: MFPB solutions.

Each MFPB solution has a definite state of polarization characterized by one of the following field components being zero everywhere in the beam, for all degrees of (strong and week) focusing: $E_x$ or $E_y$ or $B_x$ or $B_y$. Of course any linear superposition of MFPB solutions is also an exact solution of Maxwell's equations in vacuum.

We obtain each complex electric and magnetic field solution from a complex vector potential that satisfies

$$\vec{A}(\vec{r},t) = f_0 e^{-i\omega t} \, curl\left[\hat{e} \, sin\, cR\right] \quad \text{ergs/C} \tag{1}$$

where $f_0$ ergs C$^{-1}$ cm is a complex constant whose magnitude will determine the powers of the various real beam-like field solutions. Here and elsewhere sinc$R$ will always mean $R^{-1}\sin R$, while in all other contexts, $c$ will always mean the velocity of light in vacuum. The dimensionless function $R$ is defined by

$$R(x,y,z,q) \equiv \frac{\omega}{c}\sqrt{(z+iq)^2 + x^2 + y^2} \, . \tag{2}$$

We will use Gaussian-cgs units throughout, and symbolize the Gaussian unit of electric charge by C. Here $\hat{e}$ can be any normalized superposition of the Cartesian unit vectors $\hat{x}$ and $\hat{y}$ that are perpendicular to $\hat{z}$. Each solution has an evident focal plane which we place at $z = 0$. The real, positive, parameter $Q \equiv \omega q/c$ will be found to be roughly analogous to the "$f$-number" of the beam. We will show in the appendix A that, if the boundary conditions on the fields require that the beam carries no net energy into a very large sphere encircling the focal region, and if the complex vector potential is an analytic function everywhere inside the sphere (no poles), then

$$\nabla^2 \, sin\, cR = -\frac{\omega}{c} sin\, cR \, . \tag{3}$$

With (2) and (3) in equation (1) we see that $\vec{A}(\vec{r},t)$ satisfies the wave equation for velocity $c$. Because of the *curl* in (1), $\vec{A}(\vec{r},t)$ is in the Coulomb gauge. We derive explicit formulae for the five nonzero complex field components in Section **2** assuming $\hat{e} = \hat{y}$. We also examine the pronounced electric and magnetic field "bubbles" and "vacancies" (often odd in $x$) that occur in strongly-focused ($Q \ll 1$) examples of (1). We will estimate the maximum magnitudes of the electric and magnetic fields near the focus. In section **3** we derived the power $P$ and the beam intensity $I_z(x, y, z)$ for each solution. There we explore nearly-circular beam cross-sections of weakly-focused solutions. In section **4** we find that the weakly-focused MFPB solution diverges at the same angle in the far field as does the (approximate) Gaussian beam solution having the same area at $z = 0$ (at least to measurable accuracies). However, for $q$ equal to $c/\omega$ or smaller (i.e., $Q <$ 1), the errors of the corresponding (approximate) TEM$_{0,0}$ Gaussian mode solution make it of little use.

Other properties of these monochromatic finite-power beam (MFPB) solutions of Maxwell's equations in vacuum emerge.

## 2. Complex monochromatic finite-power beam (MFPB) solutions of Maxwell's Equations in vacuum.

For $\hat{e} = \hat{y}$ in (1), the electric and magnetic fields of an MFPB solution are

$$E_x = -i\frac{\omega}{c} f_0 e^{-i\omega t}\left(\frac{\partial \, sin\, cR}{\partial z}\right) \quad (4)$$

$$E_y = 0 \quad (5)$$

$$E_z = i\frac{\omega}{c} f_0 e^{-i\omega t}\left(\frac{\partial \, sin\, cR}{\partial x}\right) \quad (6)$$

$$B_x = f_0 e^{-i\omega t}\hat{x}\cdot grad\left(\frac{\partial \, sin\, cR}{\partial y}\right) \quad (7)$$

$$B_y = f_0 e^{-i\omega t}\left(\hat{y}\cdot grad\frac{\partial \, sin\, cR}{\partial y} - \nabla^2 \, sin\, cR\right) \quad (8)$$

$$B_z = f_0 e^{-i\omega t}\hat{z}\cdot grad\frac{\partial \, sin\, cR}{\partial y} \quad (9)$$

Keep in mind that the real parts of (4) - (9) describe one physical MFPB solution, while the imaginary parts describe another. Also, because these fields are in vacuum, substituting **E** for **B** and –**B** for **E** also gives new physical MFPB solutions. In addition, using $\hat{x}$ for $\hat{e}$ in (1) gives more independent MFPB solutions with beam axis along $\hat{z}$; these are of course just copies of the solutions above rotated by 90 degrees about the z axis. In Fig. 1 we show a projection of the magnetic field components (7) and (8) at the focal plane (z = 0) of a 633 nm beam for Q = 0.1 (strong focusing).

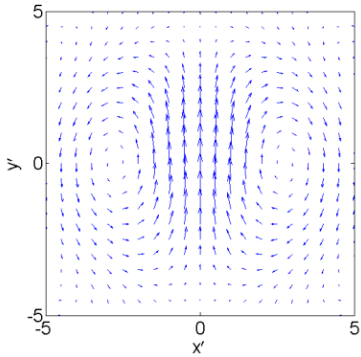

**Fig. 1**. A projection of magnetic field vector on the focal plane (z = 0) of a 633 nm He-Ne laser for Q = 0.1 (strong focusing).

## 3. The Poynting vector $\vec{I}(\vec{r},\omega)$ of a MFPB solution of Maxwell's equations in vacuum.

The time-averaged energy-flux density $\vec{I}(\vec{r},\omega)$ of a typical real (physical) MFPB solution (of angular frequency $\omega$) as derived from (4)-(9) is

$$\vec{I}(\vec{r},\omega) = \frac{c}{4\pi}\left\langle[Re\,\mathbf{E}]\times[Re\,\mathbf{B}]\right\rangle \text{ ergs cm}^{-2}\text{ s}^{-1}, \quad (10)$$

where < > indicates the time-average over many optical cycles. Using (10) to calculate the component $I_z$ of energy flux $\vec{I}(\vec{r},\omega)$ that is parallel to the beam direction $\hat{z}$, we find

$$I_z = \frac{\omega|f_0|^2}{8\pi}Im\left(\left(\frac{\partial^2 \, sin\, cR}{\partial y^2}+\frac{\omega^2}{c^2}sin\, cR\right)\left(\frac{\partial \, sin\, cR^*}{\partial z}\right)\right) \quad (11)$$

which is odd in z and so has a focal plane at z = 0. Note that we would have found this same result (11) had we used the imaginary part of the complex fields in (10).

Plots of (11) for the on-axis beam intensity for $Q = 10^{-5}, 10^{-3}, 10^{-1}, 1, 5$ and 10 are shown in Fig. 2 where they show full widths at half maximum, 5.8, 5.8, 5.8, 6.3, 11 and 23 respectively (in units of $c/\omega$). Here we see that for the regime of Q where the Gaussian beam approximation fails (Q < 1), the on-axis intensity has a FWHM of very close to one vacuum wavelength, i.e., the focal region is ~ one wavelength thick.

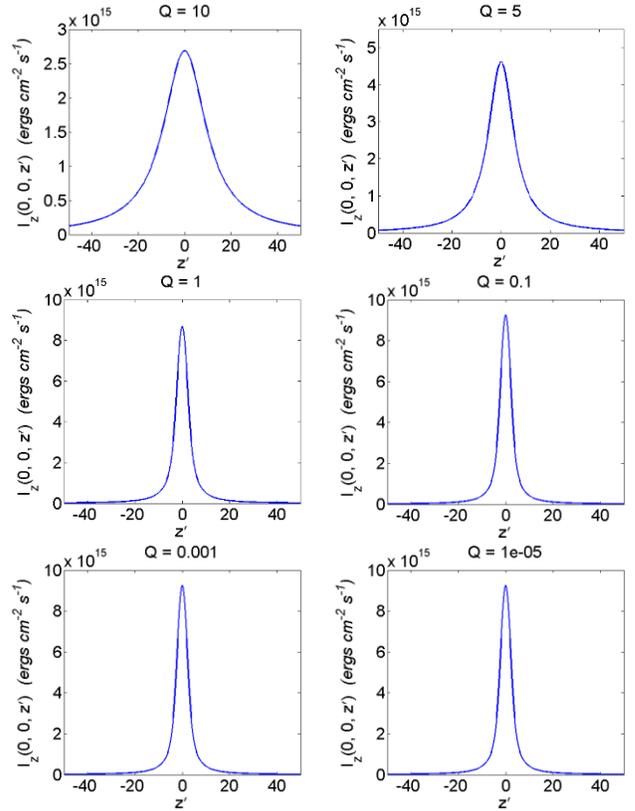

**Fig. 2**. Energy flux density component $I_z(0, 0, z')$ of a 1 Watt 633 nm He-Ne laser for $Q = 10, 5, 1, 10^{-1}, 10^{-3}, 10^{-5}$.

2D plots in Fig. 3 show how the beam intensity of a weekly-focused (Q = 10) beam varies in the planes z = 0, 10, and 20 (in units of $c/\omega$). Fig. 4 shows similar plots for Q = 1 where the Gaussian beam solution is beginning to fail.

The total power P of this beam is

$$P = \iint_{-\infty}^{+\infty} I_z dxdy \text{ ergs/sec} \quad (12)$$

For the numerical analyses below, we use real dimensionless coordinates $x' = x\omega/c$, $y' = y\omega/c$, $z' = z\omega/c$ with the dimensionless parameter $Q = q\omega/c$, so that (12) can be rewritten

$$P(f_0,\omega,Q) \quad (13)$$
$$= P_0\iint_{-\infty}^{+\infty} dx'dy'\, Im\left(\left(\frac{\partial^2 \, sin\, cL}{\partial y'^2}+\frac{\omega^2}{c^2}sin\, cL\right)\left(\frac{\partial \, sin\, cL^*}{\partial z'}\right)\right)$$

where $P_0 = (\omega |f_0|)^2/(8\pi c)$ ergs/s. Here, $L(x', y', z', Q) \equiv [(z' + iQ)^2 + x'^2 + y'^2]^{-1/2}$ is dimensionless. We have not been able to integrate (13) analytically, but our many dozens of numerical evaluations of (13), for a wide range of $Q$, show $P$ in (13) to be independent of $z$ (or $z'$). In Fig. 5 we show numerical results for $P/P_0$ from (13), as well as approximate formulae that fit these results for a wide range of $Q$.

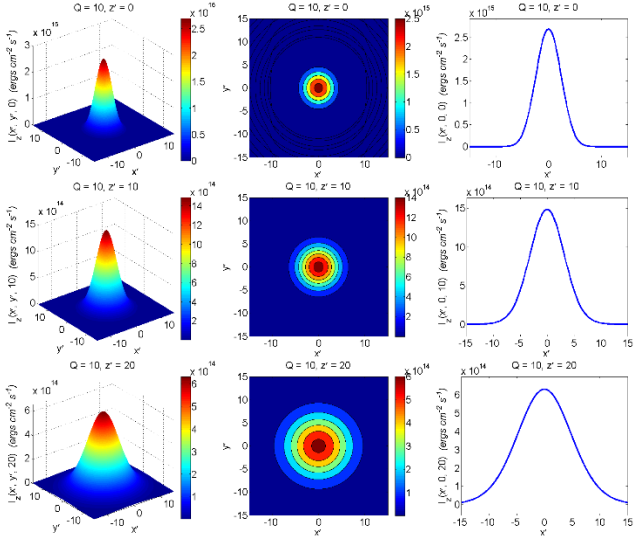

**Fig. 3**. Cross-section of energy flux density component $I_z(x', y', z')$ for a 1 Watt 633 nm He-Ne laser at planes $z' = 0, 10, 20$ for $Q = 10$.

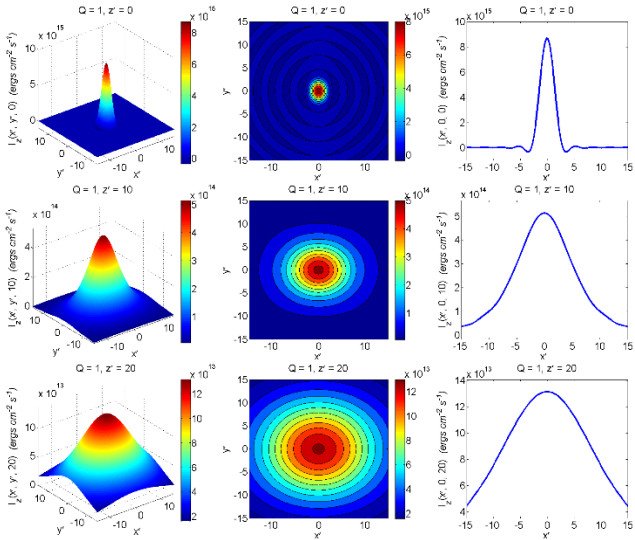

**Fig. 4**. Cross-section of energy flux density component $I_z(x', y', z')$ for a 1 Watt 633 nm He-Ne laser at planes $z' = 0, 10, 20$ for $Q = 1$.

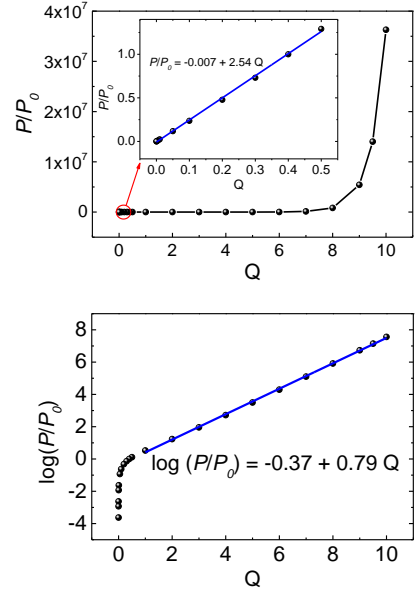

**Fig. 5**. (a) Normalized beam power $P/P_0$ as a function of $Q$. Inset: clarification of $P/P_0$ for low $Q$ ($Q < 0.5$). It can be fitted by $P/P_0 = -0.007 + 2.54Q$. (b) $\log(P/P_0)$ as a function of $Q$. It can be fitted by $\log(P/P_0) = -0.37 + 0.79Q$.

## 4. Comparing the MFPB solution to the (approximate) Gaussian beam solution.

In order to compare different exact or approximate solutions for the electric and magnetic fields of a monochromatic electromagnetic beam proportional to $\exp(-i\omega t)$, we compare plots of an "effective beam area" $S(z)$ along the beam axis ($z$), using the definition

$$S(z) = \frac{\text{total beam power } P}{\text{beam intensity } I_z(0,0,z) \text{ on axis}}. \quad (14)$$

The time-averaged intensity of the (approximate) $TEM_{0,0}$ Gaussian beam solution of Maxwell's equations, focused at $z = 0$ in vacuum, is

$$I_G(x, y, z, A) = I_A \exp\left(-\frac{(x^2 + y^2)A\omega}{(A^2 + z^2)c}\right) \text{ ergs cm}^{-2} \text{ s}^{-1}, \quad (15)$$

where $A$ is the "length" of the focal region. In this case (14) gives for the effective area:

$$S_G(z) = S_G(0)\left[1 + \left(\frac{\pi c z}{\omega S_G(0)}\right)^2\right] \quad (16)$$

which we compare to the effective area $S_{MFPB}(z)$ of a corresponding exact solution in our Fig. 6 below. The small difference between the two curves when $Q = 10$ may be partly attributed to the fact that the MFPB intensity in (11) falls off as $x^{-5}$ far from the beam axis while the intensity (15) falls off in a Gaussian manner. However, each MFPB solution derived from (1) gives another MFPB solution if the right hand side if (1) is first differentiated any number of times with respect to the parameter $q$; members of this extended family fall off as a higher power of $x$ than 5 far from the beam axis.

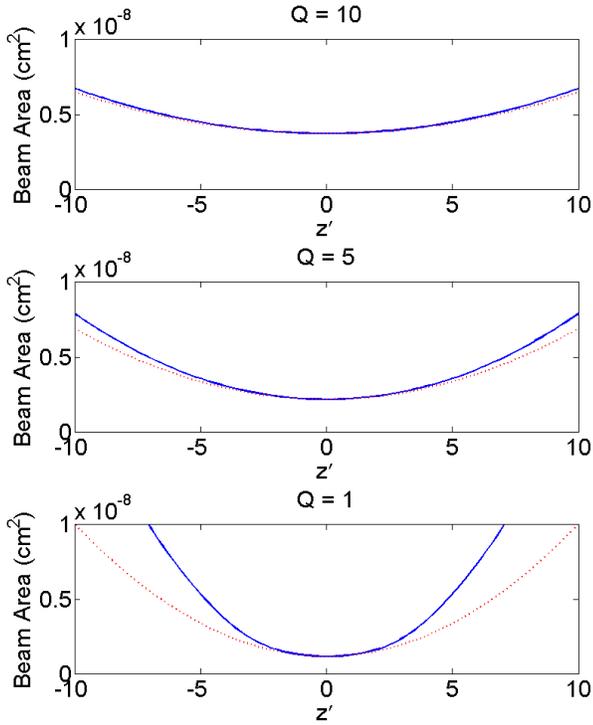

**Fig. 6**. Effective beam area of MFPB solution $S_{MFPB}(z)$ (solid blue lines) for $Q = 10, 5, 1$, compared with that of Gaussian beam solution $S_G(z)$ with the same beam area at $z = 0$.

## 5. Conclusion.

We have derived and examined a new set of exact monochromatic finite-power solutions of Maxwell's equations in vacuum. Each solution transmits a definite finite total energy flux $P$ through every plane that is perpendicular to the beam axis. The real, positive, parameter $Q$ in the solution is found to be roughly analogous to the "$f$-number" of the beam. For low Q (strong focusing) condition, power of the beam increases linearly with Q; for high Q (weakly-focusing) condition, power of the beam increases exponentially with Q. We compared the effective beam area of our solution with that of the Gaussian beam approximate solution. The beam area of our exact solution at the weakly focused condition is similar to that of the Gaussian beam, while in strong focusing condition, Gaussian solution starts to fail.

## Appendix A.

We can express $\exp(-i\omega t)\mathrm{sinc} R$ in (1) by the integral

$$e^{-i\omega t} \operatorname{sinc} cR \equiv -\int_{-\infty}^{+\infty} dt_0 \frac{e^{i\omega t_0} e^{-\frac{\omega q_+}{c}}}{2\pi\omega(t - t_0 - t_+)(t - t_0 - t_-)} \quad (17)$$

where $ct_\pm \equiv iq_+ \pm \left[(z + iq_-)^2 + x^2 + y^2\right]^{1/2}$ with $q_\pm = (q_1 \pm q_2)/2$, and $q_2 > q_1$ is real. [1, 2, 3] In terms of these parameters, the real positive parameter $q$ in (2) equals $q_-$. Using the fact that the two poles in the integrand in (17) were shown to lie in the upper half of the complex $t_o$ plane in references [2] and [3], the integral on the right hand side of (17) is readily performed by contour integration. Also, the integrand in the right hand side of (17) was shown in [2] and [3] to be an exact complex solution of the scalar wave function of $(x,y,z,t)$ with velocity $c$. So, must also be the left-hand side of (17), thus proving (3).